\documentclass{article}

\usepackage[top=1in, bottom=1in, left=1in, right=1in]{geometry}

\usepackage{microtype}
\usepackage{graphicx}
\usepackage[numbers]{natbib}
\usepackage{booktabs} 

\usepackage{hyperref}
\usepackage{color}
\usepackage{wrapfig}

\usepackage{amsmath}
\usepackage{amssymb}
\usepackage{parskip}
\usepackage{algorithm}
\usepackage{algorithmicx} 
\usepackage{algpseudocode} 
\usepackage{xcolor}
\usepackage{graphicx}
\usepackage{amsthm}
\usepackage{float}
\usepackage{physics}
\usepackage{subfig}

\newcommand{\E}{\mathbb{E}}

\newcommand{\hy}{\hat{y}}
\newcommand{\der}{\delta_{reward}}
\newcommand{\dep}{\delta_{penalty}}
\newcommand{\dc}{\text{disc}}

\newcommand\Z{\mathbb Z}

\newcommand\R{\mathbb R}

\newtheorem{theorem}{Theorem}[section]

\newtheorem{remark}[theorem]{Remark}

\newtheorem{claim}[theorem]{Claim}

\def\h_#1{\hat{#1}}
\def\wh_#1{\widehat{#1}}

\newcommand\set[1]{\left\{#1\right\}} 

\usepackage[textsize=scriptsize,backgroundcolor=green]{todonotes}

\newenvironment{varalgorithm}[1]
  {\algorithm\renewcommand{\thealgorithm}{#1}}
  {\endalgorithm}

\AtBeginDocument{%
  \providecommand\BibTeX{{%
    \normalfont B\kern-0.5em{\scshape i\kern-0.25em b}\kern-0.8em\TeX}}}

\usepackage{cleveref}
\crefformat{footnote}{#2\footnotemark[#1]#3}

\title{\bf Designing Closed Human-in-the-loop Deferral Pipelines}
\author{Vijay Keswani\footnote{Work done while at Amazon.} \\ Yale University \and Matthew Lease \\ University of Texas at Austin\\ Amazon AWS AI \and Krishnaram Kenthapadi \\ Fiddler AI}
\date{}

\begin{document}
\maketitle

\begin{abstract}
In hybrid human-machine deferral frameworks, a classifier can defer uncertain cases to human decision-makers (who are often themselves fallible). Prior work on simultaneous training of such classifier and deferral models has typically assumed access to an oracle during training to obtain true class labels for training samples, but in practice there often is no such oracle. In contrast, we consider a “closed” decision-making pipeline in which the same fallible human decision-makers used in deferral also provide training labels. How can imperfect and biased human expert labels be used to train a fair and accurate deferral framework? Our key insight is that by exploiting weak prior information, we can match experts to input examples to ensure fairness and accuracy of the resulting deferral framework, even when imperfect and biased experts are used in place of ground truth labels.  The efficacy of our approach is shown both by theoretical analysis and by evaluation on two tasks.
    
\end{abstract}



\section{Introduction}

Human-in-the-loop frameworks create an opportunity to improve classification accuracy beyond what is possible with fully-automated classification and prediction algorithms. Example applications range
from decision-making tasks that can be handled by crowd workers (e.g., content moderation \cite{davidson2017automated}) to tasks that require prior training for the humans to assist a machine (e.g., healthcare frameworks or risk assessment). 
%
Prior work has emphasized \textit{deferring} to human experts when the automated prediction confidence is low (e.g., in healthcare \cite{kieseberg2016trust})
or to include humans in auditing the automated predictions to address issues in the training of these tools (e.g., in child maltreatment hotline screening \cite{chouldechova2018case}). 

Designing effective human-in-the-loop frameworks is challenging, and prior work has proposed a variety of designs for how 
automated classifier and human experts (or a combination of both) can be best utilized to make optimal decisions in an input-specific manner \cite{keswani2021towards, mozannar2020consistent, madras2018predict}. Key challenges in designing such frameworks include:
(i) choosing appropriate model and training mechanism; (ii) addressing human and data biases; and (iii) ensuring robustness of the framework given often limited and biased training data. Of these three challenges, the third has received relatively less attention.

Consider the following example. 
Suppose a company wants to construct a semi-automated hiring pipeline \cite{schumann2020we},
and provides a small number of their employees to assist with this hiring process; the accuracy of an employee depends crucially on their fields of expertise and their implicit biases.
The company’s goal is to partly automate the hiring process and train a classifier to make the decision for most applicants in a fair and accurate manner. 
If the classifier is not confident for an applicant,
the decision should be deferred to an appropriate employee or a group of employees.  However, because this task may be highly company-specific, no suitable training dataset may exist to train the pipeline \cite{pan2021adoption}. To address this, the company might ask the same employees to label the initial input samples in order to train the framework in an online manner, with the goal of gradually shifting greater decision-making load onto the classifier after it has been appropriately trained.
%
We call such a pipeline, where the human experts available for deferral are also employed to obtain ground truth information, a \textit{closed deferral pipeline}.
%
{
Importantly, this closed pipeline combines two important modules of a normal training pipeline: the data annotation using human expert(s) module and the optimization module (usually treated separately in prior work), while ensuring that contentious future inputs can still be deferred to the same human experts.
}


A key challenge 
with such a pipeline is how to address the inaccuracies and biases of the human experts involved.
While one can collect and aggregate the decisions of the human experts 
to train the classifier and deferrer using known algorithms \cite{keswani2021towards, mozannar2020consistent}, 
large inaccuracies or biases in initial training iterations can result in a slow/non-converging training process. Furthermore, we must manage and mitigate the risk that the humans-in-the-loop may be biased, 
in which case acceptance of group majority decisions 
as the ground truth could further amplify biases 
in the final predictions \cite{mehrabi2019survey}.
%
{Particularly problematic are the human biases that originate from a lack of background, training, and/or implicit prejudices for a given task, resulting in reduced individual expert and overall framework performance for certain demographic groups. For example, the bias could be a result of human experts having high accuracy for certain specific domains while having lower accuracies for other domains
(as observed in many settings, including recruitment \cite{bertrand2004emily} and healthcare \cite{raghu2019direct}). Even in the multiple-experts setting, lack of heterogeneity within the group of experts can result in low performance for diverse input samples (as observed in case of content moderation \cite{gorwa2020algorithmic}).
}





{\bf Contributions}. 
%
%
{
Our primary contributions are the study of a closed deferral pipeline and effective algorithms to train such a pipeline using noisy labels from human experts available for deferral.
Through theoretical analysis and empirical evaluation,
we show that our approach yields an accurate and unbiased pipeline.}
%
%
To train the closed deferral pipeline using noisy human labels, we propose an online learning framework (\S\ref{sec:training_overview}). 
While known training algorithms \cite{keswani2021towards,mozannar2020consistent} can be used to directly train the closed pipeline using noisy labels, we find that using these algorithms can lead to inaccurate and biased pipelines, e.g.,  
%
when the majority of humans-in-the-loop are biased against a particular demographic group.
To address this, we propose to use 
prior information about human experts' similarity with input samples (e.g., by matching each expert's background/demographics to the input categories)
to construct an effective initial deferrer 
(\S\ref{sec:sim_function}). 
Using this initial deferrer, we can obtain relatively accurate class labels for initial inputs
and bootstrap an appropriate training process.
We present two algorithms for training this framework;
%
our first algorithm directly uses the similarity information to obtain an initial deferrer (\S\ref{sec:preproc_algorithm}), while our second algorithm provides a smoother transition from the initial deferrer to the deferrer learnt during training (\S\ref{sec:ee_algorithm}).
Empirical analyses 
over multiple datasets 
show that our proposed algorithms can tackle the inaccuracies and biases of data and  experts
(\S\ref{sec:experiments}).



\begin{figure}
    \centering
    \includegraphics[width=0.9\linewidth]{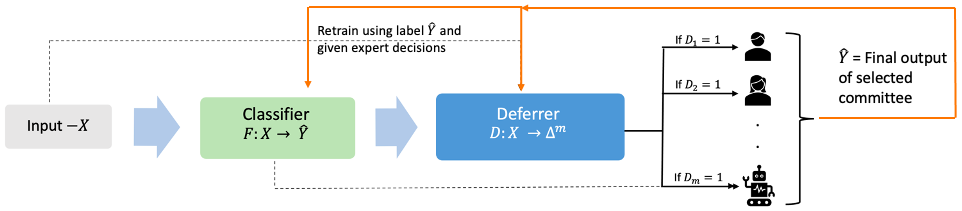}
    \vspace{-1em}
    \caption{A closed human-in-the-loop deferral pipeline.}
    \label{fig:process}
    \vspace{-1em}
\end{figure}


\section{Model and Algorithm}
%
We focus on binary classification.
Each sample in the domain contains an $n$-dimensional feature vector of the sample, denoted by $x \in \mathcal{X} \subseteq \R^n$.
$\mathcal{X}$ is the space of features available for use by the automated classifier. Human experts, on the other hand, can extract additional context about input samples, denoted by $w \in \mathcal{W}$, capturing 
%
specific information only available/interpretable by people (e.g., via prior experience or task-specific training).
We will refer to $x$ as the \textit{default attributes} and $w$ as the \textit{additional attributes}.
These attributes are used to predict a binary class label $y \in \set{0,1}$ for each sample.
%
Every sample also has a group attribute $z \in \mathcal{Z}$ associated with it (this can be either a protected attribute, e.g., gender or race, or any other task-specific interpretable categorization of feature space, e.g., assignment of inputs into interpretable sub-fields).

Our goal is to design: (i) a classifier to automatically (and accurately) predict most input samples; and (ii) a deferrer that allocates the input samples to appropriate human experts when the classifier has low confidence.
To that end, we have $(m-1)$ human experts $e_1, \dots, e_{m-1} : \mathcal{X} \times \mathcal{W}\rightarrow \set{0,1}$ available for deferral.
We use $v \in \mathcal{V}$ to denote the trainable parameters of a classifier and $f_v : \mathcal{X} \rightarrow [0,1]$ to denote the the classifier itself. It takes as input the default attributes of an input sample and returns the probability of class label being 1 for this sample.
For any given input with default features $x$ and additional features $w$, let $y_e$ denote the vector of all predictions, i.e., $ y_e(x,w):= [e_1(x,w), \ldots,  e_{m-1}(x,w), f_v(x)].$

For the sake of brevity, we will often refer to the classifier as the $m$-th expert.
Each human expert will also have additional input-specific costs associated with their predictions, denoted by 
$c_{e_j}: \mathcal{X} \rightarrow \R$ for expert $e_j$; these functions capture the time and resources that need to be spent by the expert to make the prediction for any given input.
%
The concept of different human experts having different costs is grounded in the recent tools that employ task-specific humans for label elicitation.
For example, platforms such as Upwork\footnote{\url{https://www.upwork.com/}} \cite{ipeirotis2012mechanical} and Amazon SageMaker Ground Truth\footnote{\url{https://aws.amazon.com/sagemaker/data-labeling/}} allow clients to employ the platform's experts (or freelancers) for their posted jobs. Such experts often have more expertise or experience (with correspondingly higher price) than generalist workers on Mechanical Turk \cite{barr2006ai}. 

The deferrer takes as input the default features and returns a distribution over the experts (including the classifier). 
We use $u \in \mathcal{U}$ to denote the trainable parameters of a deferrer\footnote{$\Delta^m$ denotes the $m$-dimensional simplex, i.e., for all $d \in \Delta^m$, $d_i \in [0,1]$ for all $i \in \set{1, \dots, m}$ and $d^\top \mathbf{1} = 1$.} 
 $D_u: \mathcal{X} \rightarrow \Delta^m$.
Each element $D_u(x)_i$ denotes the weight assigned to the prediction of expert $i$; the weight assigned to an expert should ideally take into account both the expert's accuracy and the cost of obtaining their prediction.
When the parameters are clear from context or not relevant to the associated discussion, we will drop the subtexts and use $D$ to denote $D_u$ and $f$ to denote $f_v$.

\begin{remark}
Since only the human experts can access the additional feature space $\mathcal{W}$, we cannot learn models to simulate human expert predictions and, correspondingly, prior ensemble learning methods \cite{oza2008classifier} cannot be directly employed for this problem.
However, using the available default features from $\mathcal{X}$, we can learn whether an expert will be correct for any given input. The deferrer indeed aims to accomplish this task; however, instead of creating $(m-1)$ error prediction models (one for each expert), we learn a single model $D$ so that the feature space is appropriately partitioned amongst the available experts.
\end{remark}

%

\textbf{Aggregation.} The final prediction of the pipeline for any input sample $(x,w)$ can be computed by combining the deferrer output $D(x)$ and expert predictions $y_e(x,w)$ in different ways. 
%
For example, the deferrer can choose $k \in \set{1, \dots, m}$ experts to make the final decision by sampling from the distribution $D(x)$ $k$ times (with replacement) to form a sub-committee of experts; the final decision is the majority decision of the sub-committee.
%
If $k=1$, the pipeline only defers to a single expert (i.e., either the classifier or a human).
%
If all experts are selected, then the framework simply returns $\mathbf{1}[D(x)^\top y_e(x,w) > 0.5]$. However, due to possibly different expert costs, it is unlikely that such an aggregation procedure would be useful for real-world applications.
%
%
Choosing an appropriate $k$
will ensure that different expert costs are taken into account and the consulted set of experts is not too expensive.
%


\textbf{Online learning.} We assume that the pipeline is trained in an online manner; the input samples arrive in a data-stream and after making each prediction, the sample can be used to re-train the classifier and deferrer (or used for batch training; see {\bf Figure~\ref{fig:process}}).
%
%
%
Learning the framework in an online manner also increases its applicability in real-world applications \cite{fontenla2013online, awerbuch2008online}.
In the absence of training data, one can still deploy this pipeline in practice, using the available human experts to make decisions during the initial iterations, and simultaneously training the classifier to steadily pick-up the decision-making load \cite{laput2015zensors} and use human experts only for deferral in later iterations.
%
%
However, as we will see in the following sections, this process can lead to problematic predictions when starting with an improper deferrer configuration. The algorithms we propose are thus designed address this problem.

\subsection{Training a deferral framework} \label{sec:training_overview}
An ideal training process for a deferral framework learns the partition of the feature space and assigns the experts to only those partitions where they are expected to be accurate.
Prior training approaches use available labeled datasets and various optimization procedures to learn a deferrer that simulates such a partition \cite{keswani2021towards}.
%
In this section, we first summarize these training procedures (that assume access to true training labels) on a high level and then later discuss extensions of these procedures for training closed pipelines.
%

\textbf{Training deferral when ground truth labels are available.} 
%
%
%
%
%
Suppose that the training input samples arrive in an online manner and each (labeled) input sample is used to train the framework.
At the current iteration of the online training process, assume we observe an input sample with default attributes $x$, additional attributes $w$, and protected attribute $z$. Let $y$ denote the true class label of this attribute, 
%
${e_i}(x,w) \in \{ 0,1\}$ denote the output of expert $i$ for input $(x,w)$, and $D(x)$ denote the deferrer output.
A general training algorithm for the deferrer will update it to reward the correct experts (for whom $e_i(x,w) = y$) and penalize the incorrect experts: 
%
\begin{equation}
D(x)_i = \begin{cases} D(x)_i + \delta_{reward}(x,y),  &\text{ if } {e_i} = y\\ D(x)_i - \delta_{penalty}(x,y),  &\text{ if } {e_i} \neq y \end{cases} 
\label{updates-True} 
\end{equation}
where $\delta_{reward}(\cdot), \delta_{penalty}(\cdot)$ are input-specific updates and will be chosen in manner that ensures that  weights $\sum D(x) = 1$.
This training process rewards and penalizes the experts based on their prediction appropriately, leading to deferrer parameter updates that simulate these rewards/penalties.
Concretely, one can show that prior training approaches for task allocation or deferral models follow this standard approach of reward/penalty updates.
%

\textit{Prior deferral models.} First, we can show that previously proposed training for deferral models for both multi-expert  \cite{keswani2021towards} and single-expert \cite{mozannar2020consistent} settings follow the above update process.
For training input $(x,w,y)$, classifier parameters $v$, and deferrer parameters $u$, these frameworks calculate the (probabilistic) output prediction as follows:
$ \hat{y}_{u,v} := \sigma (D_u(x)^\top y_{e}(x,w)),$ where $\sigma(x) := \exp(x)/(\exp(x) + \exp(1-x))$ and
\begin{eqnarray*}   
y_e(x,w) =[e_1(x,w), \ldots,  e_{m-1}(x,w), f_v(x)]
\end{eqnarray*}
As mentioned earlier, the cost associated with deferral to expert $e_j$ is denoted by the function $c_{e_j}$. 
Let $c(x) := [c_{e_1}(x), \dots, c_{e_{m-1}}(x), 0]$.
%
%
The classifier is trained to minimize a standard predictive loss function, for example, logistic log-loss:  
$$\mathcal{L}_f(u) :=  \E_{x,w,y} \left[ - y \log(f_u(x)) + (1-y)\log (1 - f_u(x))) \right]$$
The deferrer, on the other hand, is trained to minimize the following modified regularized log-loss function: $\mathcal{L}_D(u, v; \lambda) :=\E_{x,w,y}\left[ -y \log(\hat{y}_{u,v}) - (1-y)\log (1 - \hat{y}_{u,v}))  + \lambda \cdot D_u(x)^\top c(x) \right]$
where $\lambda > 0$ is the cost-hyperparameter.
The classifier and deferrer can then be simultaneously trained by combining loss functions:
\[\mathcal{L}(u,v; \lambda, \alpha) = \mathcal{L}_f(u) + \alpha \mathcal{L}_D(u, v; \lambda),\]
where hyperparameter $\alpha$ controls the sequence of classifier and deferrer training.
%
The expected loss function can be empirically computed by taking the mean over a batch of training samples, with optimization performed via gradient descent (see the \textbf{UpdateModel Algorithm} in the Appendix).
%
%
Crucially, from Theorem 2.3 of \cite{keswani2021towards}, the gradient updates for this loss function can be seen to reward the correct experts and penalize the incorrect experts. Hence, functionally, this algorithm has a similar structure as {\bf Equation~\ref{updates-True}}. 

\textit{Multiplicative weights update (MWU) algorithm.} This algorithm estimates a weight distribution over the available experts based on their predictions in order to accurately simulate their performance. 
In particular, weighted majority MWU \cite{arora2012multiplicative} works in an online manner and follows a similar training approach as Eqn~\ref{updates-True}. At iteration $t$, suppose $(w_{e_1}^{(t)}, \cdots, w_{e_m}^{(t)})$ are the weights assigned to the $m$ experts. For a pre-defined $\eta \in [0,0.5]$, if expert $j$ predicts incorrectly in iteration $t$, then its weight is decreased by a factor of $(1-\eta)$; i.e., for an incorrect expert $e_j$, $w_{e_j}^{(t+1)} = (1-\eta) w_{e_j}^{(t)}$. 
%
After normalization, this amounts to rewarding the weights of correct experts and penalizing the weights of incorrect experts.

Two important differences between the setting tackled by MWU and the deferral setting are: (i) the presence of a classifier that needs to be simultaneously trained; and (ii) the necessity of constructing an input-specific task-allocation policy.
The deferral training approach specified above addresses both of these differences.
%

\textit{Contextual multi-arm bandits (CMAB).}
Treating available user information as ``context'' (default features in our setting), one could apply a CMAB framework to find context-specific actions (from a set of possible actions) to maximize the total reward. 
Popular CMAB algorithms (e.g.,  \cite{lu2010contextual}) assume the existence of a meaningful partition of the feature space such that each input context can be assigned to a specific partition, over which a standard MAB algorithm can be run.
Considering MAB algorithms aim to reward/penalize human experts based on their predictions, this  amounts to making updates in a manner similar to Eqn~\ref{updates-True}. 
In comparison to a CMAB approach, (i) we adopt a general supervised learning approach whereas MAB 
approaches assume that the payoff is revealed only after making an action; and (ii) we train a classifier simultaneously whereas MAB approaches seek only to find the best available task allocation.

A common theme across all the above settings is the objective of accurate task allocation.
While prior algorithms for these settings assume the presence of perfect training class labels (or, in the MAB setting, accurate payoff information for any chosen action),
%
we consider the training performance when only aggregated noisy human labels are available, and we propose training methods that are robust to the noise.
%
Specifically, we focus here on training deferral frameworks using noisy labels; future work might also explore similar ideas to address noise in bandit/MWU settings. 

\textbf{Training using noisy aggregated class labels.} The primary setting we investigate is that in which ground truth class labels are unavailable, and 
%
we have access to only (noisy) expert labels obtained when the pipeline defers to the human experts.
For the input $(x,w)$, recall that $y_e(x,w)= [e_1(x,w), \ldots,  e_{m-1}(x,w), f(x)]$ and $D(x)$ is the deferrer output.
When ground truth class labels are unknown, one way to directly use the above training process is to treat the aggregated pipeline prediction for any input (denoted by $\hy$) as the true label.
%
For example, suppose the aggregated prediction is  $\hy := \mathbf{1}[D(x)^\top y_e(x,w) > 0.5]$ (i.e., defer to all experts). Then, the training updates can substitute $y$ with $\hat{y}$ in Equation~\ref{updates-True}:
\begin{equation}
D(x)_i = 
\begin{cases} D(x)_i + \delta_{reward}(x,\hy),  &\text{ if } {e_i} = \hy\\ D(x)_i - \delta_{penalty}(x,\hy),  &\text{ if } {e_i} \neq \hy 
\end{cases} 
\label{updates-Noisy}
\end{equation}
By substituting true class labels with aggregated labels,   existing training deferral algorithms \cite{keswani2021towards, mozannar2020consistent} can be used without any major changes. 
However, this approach also has a significant downside; 
the next section presents an example of bias in final decisions when using aggregated predictions for training. It shows that when the majority of the experts are inaccurate and biased, this training process is unable to address the shortcomings of the experts.

%
\begin{remark}
Training using labels obtained via crowdsourcing 
roughly follows a similar approach as described above, 
where ground truth labels are often obtained by collecting multiple labels per item from crowd-annotators  
for each input sample and performing aggregation \cite{Sheshadri13,zheng2017truth}
to find consensus labels.
%
Such aggregated labels may still contain noise or bias \cite{mehrabi2019survey}.
Our decision-making pipeline addresses this  by considering the input allocation to humans generating the labels to be part of the training process.
Technically, this closed pipeline combines (and jointly addresses biases in) the two important parts of a decision-making pipeline: eliciting human labels and optimizing the classifier/deferrer, usually learned separately in prior work.
\end{remark}


\textbf{Bias propagation when training using noisy labels.}
Assuming a binary protected attribute, we can show that: if (i)  the starting deferrer chooses experts randomly, and (ii) the majority of the experts are biased against or highly inaccurate with respect to a protected attribute type (e.g., the disadvantaged group), then 
the above training process leads to disparate performance with respect to the disadvantaged group.
For $\alpha> 0.5$, assume that $\alpha$ fraction of experts are biased against group $z=0$ and $(1{-}\alpha)$ fraction are biased against group $z=1$; in other words, majority of the experts are biased against one group.
Suppose that each expert behaves as follows. If the expert 
is biased against $z=j$, then they will always predict the class label correctly for input samples with $z{=}1{-}j$, but only predict correctly for samples with $z{=}j$ with probability 0.5.
%
%

Assuming no prior, the training will start with a random deferrer, i.e., $D(x)$ assigning uniform weight $1/m$ to all experts. 
%
%
%
When $k=1$, the deferrer chooses a single expert to make the final decision.
%
In this case, 
the starting accuracy for group $z=1$ elements will be $\alpha + 0.5(1-\alpha)$ and the starting accuracy for group $z=0$ elements will be $(1-\alpha) + 0.5\alpha $. Therefore, when choosing a single expert, the difference in expected accuracy for group $z=1$ elements and expected accuracy for group $z=0$ elements is $(\alpha - 0.5)$; the larger the value of $\alpha$, the greater the disparity.
Hence, the starting deferrer will be biased, and since the predicted labels are used for retraining, the bias can propagate to the learned classifier and deferrer as well.

\begin{claim} \label{clm:example}
In the above setting, the disparity between the accuracy for group $z=0$ and the accuracy for group $z=1$ does not decrease even after training using multiple Equation~\ref{updates-Noisy} steps.
\end{claim}

%
Considering the starting deferrer bias affects the training process and can lead to a biased final deferrer, it is necessary to explore ways that mitigate bias at the initial training steps or the starting deferrer itself.
The proof of the claim is presented in Appendix~\ref{sec:proofs}.

\subsection{Expert-Input similarity quantification} \label{sec:sim_function}

We seek a better training algorithm that uses expert predictions to iteratively train the classifier and deferrer while addressing the risks of bias discussed above.
As we showed in \S\ref{sec:training_overview},
starting with a random deferrer can be problematic when the majority experts are biased against a given group.
In the absence of ground truth labels, we thus require other mechanisms to calibrate the initial starting deferrer. 
In particular, we want to start with some prior information about  which expert might be accurate for each input category, and then bootstrap an accurate training process using this prior information.

\textbf{Similarity function $dSim$}. Let $dSim: \{ e_1, \dots, e_{m-1}\} \times \mathcal{Z} \rightarrow  [0,1]$ be a matching function that specifies the ``fit'' of a given expert to a given input category. 
We can incorporate this function as a prior to better initialize the training process.
%
{
Ideally, for any input category, the dSim function should assign ``large'' weights to experts accurate for this category and ``small'' weights to inaccurate experts. 
While it can difficult to preemptively infer the accuracy of an expert for an input category in real-world settings, 
%
the above similarity function occurs naturally in many applications, 
as examples below highlight.
However, the difference between the weight assigned to accurate experts and the weight assigned to inaccurate experts (or the strength of the $dSim$ function) can vary by application and context.
}

\textbf{Example 1.} In moderating social media content, if expert $e_{j_1}$ writes in the dialect of a given post being moderated and expert $e_{j_2}$ does not, then $e_{j_2}$'s decisions may be biased \cite{sap2019risk,davidson2019racial,keswani2021towards}. A  similarity function can be constructed such that $dSim(e_{j_1}, z) > dSim(e_{j_2}, z)$, where $z$ represents the dialect of the given post; e.g., similarity with first expert could be set to 1 and similarity with second expert could be set to 0.
%
Content moderation tasks may thus require 
annotators to fill out a demographic survey, and we might 
ask experts for their dialect in order to assess their match to the input samples. 

\textbf{Example 2.} $dSim$ functions can also be constructed for content moderation  even when expert demographic information and/or the dialect of the posts are unavailable.
To construct a $dSim$ function in this case, one can hand-label the class label 
of a small set of posts $T$ and then check each expert's correctness on every post of $T$.
Then, for a new post $x$, the $dSim$ value of an expert for $x$ can be computed by taking the average similarity of $x$ with the posts in $T$ where the expert was correct.
This mechanism of extracting latent information using similarity with labeled representative examples has also been employed in prior work on diversity audits \cite{keswani2021auditing}.

\textbf{Example 3.} In the setting where a company wants to construct a semi-automated recruitment pipeline \cite{schumann2020we}, the human experts could be the employees in the company itself. 
In that setting, $dSim$ would quantify the similarity between any employee's field of expertise and the applicant's desired field of employment within the company. Once again, since the company usually has data on its employees, constructing such a similarity function should be feasible.


Note that we define similarity with respect to input category $z$ rather than input sample $x$. We do so because $dSim$ is expected to be a context-dependent and interpretable function. 
Therefore, most applications where such similarities can be quantified could use $dSim$ functions over broad input categories to define similarities to the given experts.
This is also apparent from the examples provided above (demographic features in case of Example 1 on content moderation and field of expertise in case of Example 3 on recruitment).
Nevertheless, if similarity with respect to each input sample is available for any given setting, this measure could be alternately employed by treating each input sample as belonging to its own category (as in Example 2).

\begin{remark}[Disparity of $dSim$] \label{rem:example_dsim}
For the setting in
Claim~\ref{clm:example},
consider the deferrer induced by an appropriate $dSim$ function (i.e., for input $(x,z)$, we have that deferrer output $D(x)_i \propto dSim(e_i, z)$). 
%
%
Suppose 
$dSim(e_j, z) = 1$ if expert $e_j$ is unbiased for input $z$ and $\gamma$ otherwise, where $\gamma$ is any constant $\in [0,1]$.
%
Then the difference between the accuracy for group $z{=}0$ and $z{=}1$ lies in the range $\left[ \frac{\gamma}{2}, \frac{\alpha}{1-\alpha} \cdot \frac{\gamma}{2} \right]$ (proof in Appendix~\ref{sec:proofs}).
Smaller values of $\gamma$ here thus imply that $dSim$ is better able to differentiate between biased and unbiased experts for any given input.
Hence, 
the better $dSim$ is at differentiating biased and unbiased experts, the smaller the disparity will be in performance of the starting deferrer with respect to the protected attribute.
\end{remark}

\subsection{Preprocessing to find a good starting point} \label{sec:preproc_algorithm}
%

To address the issue of possible biases in training using noisy labels (i.e., Eqn~\eqref{updates-Noisy}), our key idea is to encode a prior for the initial deferrer output; this assigns weights to experts in a manner that is similar to the behavior of the $dSim$ function (extending the observation from Remark~\ref{rem:example_dsim}). 
%
%
In particular, we set initial deferrer parameters such that, for the starting deferrer $D_{u_0}$ and any input $(x,z)$ and expert $e_i$, we have that $D_{u_0}(x)_i \propto dSim(e_i, z) $.
This step can indeed be feasibly accomplished in many applications using unlabeled training samples (see Section \ref{sec:experiments} for examples).
The rest of the training process for classifier $f$  and deferrer $D$ is the same as described in  Section~\ref{sec:training_overview} and Eqn~\eqref{updates-Noisy}; i.e., for every input sample, reward the experts whose prediction matches with aggregated prediction and penalize the experts whose prediction does not match with aggregated prediction.
{To create a further robust training procedure, we can also use a batch update process; i.e., for a given integer $B$, train the deferrer and classifier after observing $B$ input samples using the batch of these $B$ samples and predictions.}
The complete details are presented in Algorithm~\ref{alg:main}.
As discussed earlier, the aggregation step (Step 6) can be executed in different ways, e.g., using all expert predictions (i.e., use $D_{u_{t-1}}(x_t)^\top y_e$) or sampling $k$ experts from distribution $D_{u_{t-1}}(x_t)$ and using their majority decision.

\begin{varalgorithm}{Strict-Matching}
\flushleft
  \caption{(using $dSim$ to encode a prior) \\{Input}: stream of input-category pairs $(x_1,z_1), (x_2,z_2)\dots$,  experts $e_1, \dots, e_{m-1}$,  initial classifier parameters $v_0$, function $dSim: \{ e_1, \dots, e_{m-1}\} \times \Z \rightarrow  [0,1]$, batch size $B$, and update algorithm \textit{UpdateModel} (Appendix~\ref{sec:update_algorithm}) with parameters $\alpha, \eta, \lambda$.}
  \begin{algorithmic}[1] 
	\State Set initial deferrer parameters $u_0$ such that for any input $x$ in category $z$, we have that $D_{u_0}(x)_i \propto dSim(e_i, z) $
	\State $S \gets \emptyset$
  	\For{$t \in \{1, 2, \dots,  \}$}
	    \State $y_e \gets [e_1(x_t), e_2(x_t), \dots, e_{m-1}(x_t), f_{v_{t-1}}(x_t)]$ 
        \State $D_{u_{t-1}}(x_t) \gets $ Deferrer output for input $x_t$
    	\State $y_t \gets $ Aggregate predictions $y_e$ using weights $D_{u_{t-1}}(x_t)$  
    	\State $S \gets S \cup \set{(x_t, y_t)}$
    	\If{$|S| == B$}
		    \State $v_t, u_t \gets$ UpdateModel$(S, v_{t-1}, u_{t-1}, \alpha, \eta, \lambda)$
		    \State $S \gets \emptyset$
		\EndIf
	\EndFor
	\State return $u_t,v_t$
\end{algorithmic}
    \label{alg:main}
\end{varalgorithm}

The main advantage of using this pre-processing approach is that we utilize the information provided by $dSim$ to start with a good deferrer (obtaining true ground truth labels in initial iterations) and proceed to use the training steps towards learning an even better deferrer.
Correspondingly, even if certain experts have low $dSim$ score for any input but high true accuracy for that input, this behaviour will be discovered and incorporated in the deferrer during later training stages.
This pre-processing to encode the starting deferrer can be considered an \textit{exploitation phase}, since we exploit prior information available via $dSim$, while the subsequent training steps can be considered an \textit{exploration phase}.
%

%



%
\subsection{Smoother exploitation-exploration} \label{sec:ee_algorithm}
To obtain a better transition from a good starting point provided by $dSim$ to the weights learnt during the training iterations, we can slowly decrease the weight assigned to the prior as the number of observed samples increases.
In other words, the deferrer employed at any iteration will be a convex combination of the deferrer encoded by the $dSim$ prior and the deferrer trained using the observed samples (and aggregated class labels). The hyperparameter used for this combination will depend on the number of observations, and the larger the number of observations, the smaller the weight assigned to the prior distribution should be.
The complete details are provided in Algorithm~\ref{alg:main_2}.

\begin{varalgorithm}{Smooth-Matching}
\flushleft
  \caption{\\
  {Input}: stream of input-category pairs $(x_1,z_1), (x_2,z_2)\dots$,  experts $e_1, \dots, e_{m-1}$,  initial classifier parameters $v_0$, initial deferrer parameters $u_0$,  $dSim$, hyperparameter $T_{d}$, batch size $B$, and update algorithm \textit{UpdateModel} (Appendix~\ref{sec:update_algorithm}) with parameters $\alpha, \eta, \lambda$.}
  \begin{algorithmic}[1] 
    \State $S \gets \emptyset$
  	\For{$t \in \{1, 2, \dots,  \}$}
	    \State $y_e \gets [e_1(x_t), e_2(x_t), \dots, e_{m-1}(x_t), f_{w_{t-1}}(x_t)]$      
	        \State $\mu \gets \frac{T_{d}}{t+T_{d}}$
	        \State $D_{dSim}(x_t) \gets [dSim(e_1, z_t), \cdots, dSim(e_{m-1},z_t), 0]$
	        \State $D_{dSim}(x_t) \gets D_{dSim}(x_t)/\text{sum}(D_{dSim}(x_t))$
	        \State $d \gets \mu \cdot D_{dSim}(x_t) + (1-\mu) \cdot D_{u_{t-1}}(x_t)$	        
    	    \State $y_t \gets $ Aggregate predictions $y_e$ using weights $d$  
        	\State $S \gets S \cup \set{(x_t, y_t)}$
        	\If{$|S| == B$}
    		    \State $v_t, u_t \gets$ UpdateModel$(S, v_{t-1}, u_{t-1}, \alpha, \eta, \lambda)$
    		    \State $S \gets \emptyset$
    		\EndIf
    	\EndFor    
	\State return $u_t,v_t$
\end{algorithmic}
    \label{alg:main_2}
\end{varalgorithm}


\ref{alg:main_2} can lead to better exploration than \ref{alg:main} in settings where
some accurate experts are not given relatively high weights by $dSim$ and need to be ``discovered'' during training and assigned higher weights.
%
This is primarily because after $T_{d}$ iterations, the weight given to prior induced by $dSim$ is relatively smaller than the the weight given to the deferrer distribution learnt during training. 
The first $T_{d}$ iterations focus on obtaining accurate ground truth information to bootstrap the training process while the subsequent iterations try to accurately partition feature space amongst the $m$ experts.

\subsection{Theoretical analysis}
For the algorithms that use $dSim$ function as a starting point for the deferrer, we can show that the final trained deferrer converges to a point that simulates the underlying accuracy functions of the experts.
Furthermore, this is true even if we start with a ``weak'' $dSim$ function.
%
%
In particular, if any expert $e_j$ has high accuracy for category $z$ then how fast the algorithm converges to a deferrer that assigns high weight to $e_j$ for category $z$ depends on how much initial weight is assigned to $e_j$ for $z$.


\begin{theorem}[Exploitation using $dSim$] \label{thm:exploitation}
For any input group $z$, say expert $e_j$ is more accurate than all the experts $e_{j'}$, for $j' \in \set{1, \dots, m} \setminus \set{j}$.
For $\beta > 0$, suppose we set $dSim$ function in a manner such that $dSim(e_j, z) - \max_{j' \in \set{1, \dots, m} \setminus \set{j}} dSim(e_{j'}, z) \geq \beta$.
Then the training algorithm that initializes the deferrer parameters with this $dSim$ function increases the weight assigned to expert $e_j$ by atleast $2\beta \delta$ amount in expectation, where $\delta \in [0,1]$ depends on the choice of $\der$ and $\dep$ values for the given input.
\end{theorem}

In other words, the smaller $\beta$ is, the weaker is the starting deferrer, and the longer it takes converge to a deferrer that assigns high weights to accurate experts.
%
Nevertheless, the theorem also shows that even if $\beta$ is small, accurate experts are positively rewarded on average so long as we start with a $dSim$ function that assigns them large weights.
The proof is presented in Appendix~\ref{sec:proofs}.
%
Next, we show that, for any input category $z$, even if there are accurate experts who are not assigned high weight by $dSim$ function, they can be ``discovered'' by the training algorithm.

\begin{theorem}[Exploration of accurate experts] \label{thm:exploration}
For any input group $z$, say expert $e_j$ has accuracy $1$.
%
Let $k$ be the size of the sub-committee sampled from the deferrer output distribution to make the final decision for the given input.
Suppose we set $dSim$ function in a manner such that $dSim(e_j, z) = \epsilon$, for some  $\epsilon \in [0,1]$, but the total weight (normalized) assigned by $dSim$ to accurate experts for group $z$ is greater than 0.5. 
Then, there is an expected positive increase in the weight of this expert if
$\epsilon > 1 - \left(1 - \frac{k}{2m} \right)^{1/k}.$
\end{theorem}
%
Hence, the training algorithm can discover accurate experts
so long as some other accurate experts are also available to infer the true labels for this input category.
However, the theorem also implies that 
%
either $\epsilon$ or $k$ needs to be large for this to happen; large $\epsilon$ would imply that this expert is given large initial weight, while large $k$ would increase the chances of this expert being sampled.


\section{Evaluation} \label{sec:experiments}
\begin{wrapfigure}{r}{0.4\textwidth}    
    \centering
    \includegraphics[width=\linewidth]{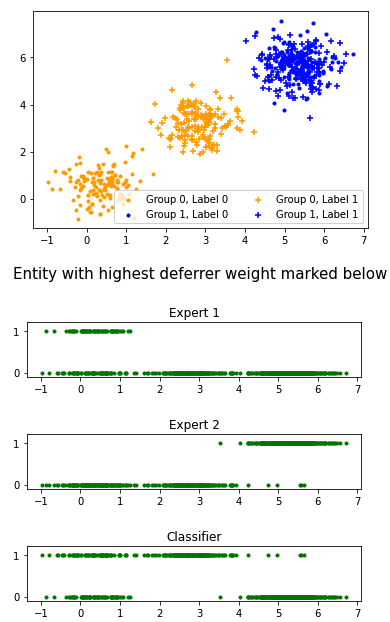}
    \caption{ Algorithm~\ref{alg:main} performance for the cluster detection task using $s{=}0.4$. The top plot shows the dataset. The lower three plots show the relative weights assigned to each expert
    (expert with highest weight assigned 1).
    All plots share the same $x$-axis and are aligned vertically; i.e., for any point in the top plot, the point with the same $x$-axis value in the bottom three plots present the weights assigned to the experts for this point.
    %
    }
    \label{fig:synthetic_results_1} 
    \vspace{-2em}
\end{wrapfigure}

    To assess the efficacy of our closed decision-making pipeline, we begin with presenting evaluation on
    a cluster detection task
    with two experts.
    Following this, we report further evaluation on a
    content moderation task
    with 40        
    experts.

\subsection{Cluster detection task}
\label{eval-syn}

%
Our first evaluation highlights the importance of learning ground truth from appropriate experts in  initial iterations and correctly partitioning the input space amongst classifier and experts.

\textbf{Dataset.} We generate 2000 examples, each represented using a 2-dimensional feature and a color attribute. The goal is to predict the class label corresponding to each example. 
We first sample $\mu \sim $Unif$[0,1]$ and construct a 2-dimensional diagonal matrix $\Sigma$, with diagonal entries sampled from Unif$[0,1]$.
We next sample 250 elements from $\mathcal{N}(\mu, \Sigma)$ (colored \textit{orange} and given class label 0), then 
sample 250 elements from $\mathcal{N}(\mu+2.5, \Sigma)$
(colored \textit{orange} and given class label 1).
Finally, 500 elements are sampled from $\mathcal{N}(\mu+5, \Sigma)$; these are colored \textit{blue} and elements are assigned class label 0 or 1 randomly.
Note that for the orange cluster, the samples with different class labels form distinct sub-clusters (i.e., can be perfectly classified).
For the blue cluster, the samples with different class labels cannot be separated using just the available 2-dimensions (see Figure~\ref{fig:synthetic_results_1}).
The best possible accuracy that can be achieved by any classifier is 0.75.
%

%
%
%

\textbf{Synthetic Experts.} 
Expert $e_1$ always predicts the class label correctly for orange samples but only 20\% of the time for blue samples. Expert $e_2$ has the opposite behavior.
Note that experts have access to the sample color, while the classifier and deferrer do not.

\textbf{Similarity function.} 
For expert $e_j$ and input sample $x$, let $z $ denote the color of sample $x$.
Then, for a pre-defined $s \in [0, 0.5]$, we set
$dSim(e_j, z) =  1-s$, if ($j=1$ \& $x$ is orange)  or ($j=2$ \& $x$ is blue), and $s$  otherwise.
Note that the higher the value of $s$, the \textit{weaker} the corresponding $dSim$ function will be.
Also $dSim(e_3, \cdot) = 0.1$ to ensure that classifier gets some small initial starting weight.
Additional details of the setup 
are provided in Appendix~\ref{sec:experiments_appendix}.

%
%
%
%
    %
    %

\begin{wrapfigure}{r}{0.5\textwidth}    
        \centering
        \includegraphics[width=\linewidth]{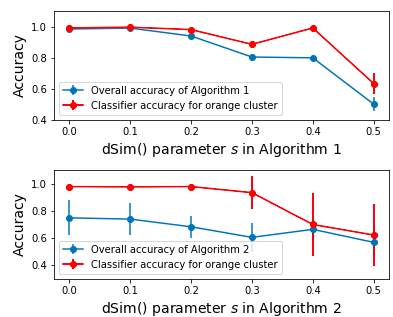}
        \caption{Algorithm~\ref{alg:main} and Algorithm~\ref{alg:main_2} on the cluster detection task for different $dSim$ functions.}
        \label{fig:acc_vs_dsim_syn} 
    \end{wrapfigure}

    %
    \textbf{Observed results.} 
    %
    \textit{Performance of Algorithm~\ref{alg:main}.} We first look at one run of Algorithm~\ref{alg:main} using the 
    $dSim$ with $s{=}0.4$. The overall accuracy achieved is 0.92. The accuracy for the orange cluster is 0.85 and the accuracy for the blue cluster is 0.99.
    \textbf{Figure~\ref{fig:synthetic_results_1}} also shows the deferrer weight distribution amongst experts $e_1, e_2$, and the classifier for different input samples.
    The blue cluster has almost-perfect accuracy since $e_2$ makes most decisions for that cluster. For the orange cluster, $e_1$ is still consulted for around 20\% input samples, but the classifier has a higher weight and makes the decision for the other 80\% input samples of this cluster.
    Importantly, note that $dSim$ with $s{=}0.4$ is not very strong.
    Nevertheless, the framework is indeed able to accurately infer ground truth labels in the initial iterations and train a highly accurate classifier for the orange cluster (classifier accuracy is 0.85).
    However, increasing the value of $s$ in the $dSim$ function leads to a corresponding decrease in accuracy.
    \textbf{Figure~\ref{fig:acc_vs_dsim_syn}} shows the performance for different $s$ values; for each $s$ parameter, we run the algorithm 10 times using a random train-test partition.
    As expected, for large values of $s$, $dSim$ does not provide any information about difference between accurate and inaccurate experts, and, hence, the accuracy in this case is pretty low.
    %


    \textit{Performance of Algorithm~\ref{alg:main_2}.} The performance of Algorithm~\ref{alg:main_2} (using $T_d = 500$), given in Figure~\ref{fig:synthetic_results_1} second plot, is relatively more varied for different $s$ values.
    The maximum overall accuracy achieved is around 0.75 (lower than that of \ref{alg:main}) and decreases with increasing $s$ value.
    %
    This is primarily because combining $dSim$ linearly with the deferrer output leads to relatively smaller updates to the deferrer parameters in each iteration.
    Nevertheless, even in this case, 
    %
    the trained classifier does have high accuracy for the orange cluster when $s{<}0.4$.

    {
    Note that when the $dSim$ function assigns equal weight to all experts for every input, then aggregated human predictions are directly used as true class labels. In our evaluation setup, using this $dSim$ function in \ref{alg:main} corresponds to using the baseline \cite{keswani2021auditing}, which in this case, has low (both overall and classifier) accuracy (~0.60).
    }
    
\subsection{Content moderation task}
\label{eval-real}
We next evaluate our algorithms on a content moderation task using an offensive language dataset and multiple synthetic experts.
By varying the number of experts assigned high $dSim$ weights, we will demonstrate the effectiveness of our algorithms in ``discovering'' experts that are accurate but assigned low $dSim$ weights.

\textbf{Dataset.} 
%
This dataset contains 25k Twitter posts that are labeled as either offensive language or not.
The posts are written by users in US and feature two primary dialects - African-American English (AAE - around 64\% of dataset) and standard English (non-AAE).
%
%
We treat dialect as the group attribute here; i.e., $z \in \set{AAE, non-AAE}$. 

\textbf{Expert design. } 
We use 40 synthetic experts for this experiment.
%
Suppose that 30 of these experts have experience mainly with the non-AAE dialect and 10 experts have experience mainly with the AAE dialect; we will call the first 30 experts \textit{non-AAE experts} and last 10 experts \textit{AAE experts}.
%
%
The behaviour of each expert is the following:
%
For the first 30 non-AAE experts, we define $p_j := 0.6 + 0.4{\cdot}j/30$, for $j \in \set{1, \dots, 30}$ and $q_j := p_j{-}0.3$.
The non-AAE expert $e_j$ predicts the label of a non-AAE post correctly with probability $p_j$ and predicts the label of an AAE post correctly with probability $q_j$.
Similarly, for the AAE experts, we define $p_j := 0.6 + 0.4{\cdot}(j{-}30)/10$, for $j \in \set{31, \dots, 40}$ and $q_j := p_j{-}0.3$.
The AAE expert $e_j$ predicts the label of an AAE post correctly with probability $p_j$ and predicts the label of a non-AAE post correctly with probability $q_j$.
Note that, in this case, 75\% of experts are biased against the AAE dialect.
We also present the variation of performance with number of experts in Appendix~\ref{sec:content_moderation_appendix}.
%


\textbf{Similarity function.} 
%
For a given integer $n_s \in \set{0, \dots, 10}$, choose an $n_s$-sized random subset of AAE experts, say $E_{AAE}$, and an $n_s$-sized random subset of non-AAE experts, say $E_{non-AAE}$.
For an input from group $z$, $dSim(e_j, z) = 1$, if $e_j \in E_z$, and 0 otherwise.
Note that, for this setting, the smaller the value of $n_s$, the weaker is the information provided by $dSim$ about expert accuracies.
In other words, by choosing small $n_s$ and observing the final behaviour of learned deferrer, we can check whether the algorithm is able to explore and find accurate experts that are not revealed by $dSim$.

\textbf{Methodology.} We use three-layer neural networks for the classifier and the deferrer. The inputs are 25-dimensional sentence-embedding of the Twitter posts generated using pre-trained GloVe models \cite{pennington2014glove,arora2016simple}.
To generate final decision for test samples, we sample $k=5$ experts from the deferrer distribution $D(\cdot)$ and return the majority decision of the selected experts as the final decision.
We perform 20 repetitions for each $n_s$, sampling random subsets $E_{AAE}$ and $E_{non-AAE}$ in each repetition (see Appendix~\ref{sec:experiments_appendix} for other details).

\begin{figure}
    \centering
    \includegraphics[width=0.9\linewidth]{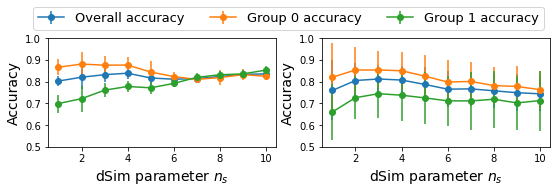}
    \subfloat[Performance of \ref{alg:main}]{\hspace{.5\linewidth}}
    \subfloat[Performance of \ref{alg:main_2}]{\hspace{.5\linewidth}} 
    \caption{Performance of our approaches for different $dSim$ function parameters on the content moderation task.}
    \label{fig:aae_acc_vs_dsim}
\end{figure}

\textbf{Results.}
%
For $n_s{=}2$, i.e., when only two experts for each dialect are assigned non-zero weights in $dSim$,
%
the overall accuracy achieved by Algorithm~\ref{alg:main} is 0.82 ($\pm$ 0.06); the accuracy for non-AAE group is 0.72 (${\pm}0.05$) and accuracy for AAE group is 0.88 ($\pm$ 0.06). 
Similarly, the overall accuracy achieved by Algorithm~\ref{alg:main_2} is 0.83 ($\pm$ 0.03); the accuracy for non-AAE group is 0.72 ($\pm$ 0.05) and accuracy for AAE group is 0.90 ($\pm$ 0.03).
In comparison, the baseline that randomly selects $k$ experts to make the decision 
has overall accuracy 0.79 ($\pm$ 0.01) with non-AAE accuracy of 0.66 ($\pm$ 0.01) and AAE accuracy of 0.87 ($\pm$ 0.01).
Similarly, $n_s=0$ corresponds to the baseline \cite{keswani2021towards}. Once again, for this baseline, there is significant disparity between AAE and non-AAE accuracy, with non-AAE accuracy around 0.70.
Hence, our approaches
lead to an improved accuracy compared to baselines, even for small non-zero $n_s$.

%
\textbf{Figure~\ref{fig:aae_acc_vs_dsim}} also shows the performance of \ref{alg:main} and \ref{alg:main_2} with respect to changing $n_s$.
%
For both algorithms, increasing $n_s$ leads to reduced disparity between accuracy for different groups.
This is because as $n_s$ increases, the likelihood of the prior $dSim$ giving high weight to accurate experts for both groups gets larger, leading to training an unbiased deferrer.
%
However, in case of \ref{alg:main_2}, the standard deviation is high and disparity is non-zero even when $n_s$ is high. This is likely due to the fact that a smooth transition from $dSim$ to the learnt deferrer leads to smaller deferrer updates in each iteration and, hence, slower convergence.
\ref{alg:main}, on the other hand, performs much better than the random committee selection baseline and \ref{alg:main_2}, reducing disparity and increasing accuracy with increasing $n_s$.

\section{Related Work}

    \textbf{Deferral frameworks.} As discussed earlier, 
    prior algorithms for deferral training assume that true class labels are available
    \cite{mozannar2020consistent,keswani2021towards,madras2018predict} 
    and are difficult to directly employ in our setting. 
    %
    Furthermore, \cite{madras2018predict, mozannar2020consistent} are also restricted in their design; they assume the availability of only a single human expert. 
    Also related is the literature on rejection learning \cite{cortes2016learning, liu2019deepgamblers}; however, here the human is not explicitly specified in the framework which leads to pipelines that are relatively less robust pipelines to individuals' biases.
    
    %
    %
    \textbf{Task allocation.} 
    %
    For task allocation, \cite{goel2019crowdsourcing} provide a framework that learns a single unbiased allocation policy, given estimates of annotators’ accuracies for different protected attribute types.
    %
    %
    The main drawback of their framework is that they use the same allocation policy for all future samples; in contrast, constructing input-specific allocation policies, like in our framework, will ensure that the domain of expertise of different human experts are more appropriately taken into account. 
    \cite{valera2018enhancing} consider the problem of task allocation in the case of risk assessment; however, their model for human predictions also has low generalizability as they assume that all human experts in the pipeline have the same prediction behavior. 
    %
    \cite{yin2020matchmaker} design budget-limited allocation policies that match the annotator preferences to task requirements 
    when experts have significant heterogeneity amongst them. 
    However, their algorithms assumes access to annotator preferences, which will not always be available and may not necessarily include the implicit biases of the experts. 
    \cite{wallace2011should} 
    provide a task allocation algorithm
    that prioritizes assigning ``difficult'' tasks to more accurate experts. 
    Unlike our approach, theirs is an offline algorithm that allocates one expert per task, which can be restrictive in many settings. 
    \cite{ho2013adaptive, karger2014budget} study a different setup where the human annotators arrive in an online fashion and the task allocation scheme chooses which input is to be deferred to the arriving annotators.
    
    %
    %
    \textbf{Learning using noisy labels.}
    \cite{nguyen2015combining, yan2011active, kajino2012convex} 
    consider the semi-supervised problem of learning training labels from multiple noisy crowd-annotators. 
    %
    %
    The main difference between our model and these papers is the presence of a deferrer, and as shown by \cite{keswani2021towards,mozannar2020consistent}, training a deferral framework is more complex than training just a classifier. 
    Furthermore, unlike their approaches, our framework utilizes the human experts available for deferral to obtain training labels as well.
    \cite{nguyen2015combining} assume access to highly-accurate experts to obtain ground truth labels for some samples. 
    In contrast, we employ measures of similarity between the experts and inputs to extract prior information about expert correctness.
    %
    {
    In the field of interactive offline learning, \cite{xu2017noise} tackle the learning problem when human experts provide pairwise comparisons between individual inputs instead of class labels; in contrast, we assume that the experts provide class label decisions for our online framework, but compare these decisions with the aggregated labels during the training phase.
    }
    %
    %
    
    %
    \textbf{Semi-supervised and reinforcement learning.} 
    %
    %
    Prior work on learning algorithms has attempted to train accurate pipelines using limited ground truth data
    \cite{chapelle2010semi}.
    Our framework, while more generic, has similarities with other weak-supervision learning approaches for certain choices of $dSim$ function.
    For example
    if $dSim$ for an expert encodes similarity between the given input sample and the previous samples accurately labeled by the expert, then using this measure for learning is related to the well-known semi-supervised self-training approach \cite{triguero2015self}.
    %
     %
     %
    In a related setting, \cite{neel2018mitigating} tackle the problem of
    exploration-exploitation tradeoff for unbiased data gathering and provide a differentially-private framework for this case. 
    \cite{ensign2018runaway} further analyze how feedback loops can reinforce negative stereotypes against minority groups. However, these works do not explicitly deal with the problem of unbiased task allocation.

    \section{{Discussion and Limitations}}
    Our proposed design of a closed pipeline and the accompanying process of training it combines the label elicitation and the model learning components of a decision-making framework. 
    A variety of future directions could be pursued to obtain a more robust pipeline.
    
    %
    \textbf{Adding/replacing experts.}
    Adding or replacing experts in our setting is relatively simple given the usage of the $dSim$ function.
    Any new expert can be assigned weight proportional to the $dSim$ value for any given input category and subsequent training steps will improve this weight 
    based on the expert's predictions.
    However, removing an expert is more difficult. Deleting its entry from the deferrer output can create issues with training since deferrer parameters had been trained assuming its presence.
    If the deleted expert is not 
    unique/necessary for obtaining accurate predictions in a particular input category, then choosing a large committee size can partially address this issue.
    %
    However, a robust way to deal with an expert's removal would be beneficial for implementing this framework in real-world applications.
    
    \textbf{Evaluation on a real-world dataset.} 
    {
    Our experimental setup demonstrates the performance of our algorithms on simplified versions of real-world scenarios.
    The cluster detection task models the cases where additional information available to experts helps them make more accurate decisions than automated classifiers for some inputs (as observed in \cite{chouldechova2018case} for maltreatment hotline screening). The content moderation task models the setting where experts with similar demographic attributes as users are better suited to judge the users’ content. 
    Nevertheless, a real-world evaluation using real human experts would be beneficial to test the performance of our framework.
    %
    This evaluation would allow us to assess the quality of $dSim$ functions that can be constructed using available data about the demographics and background of the human experts, and test performance during initial iterations in real-time.
    However, such an evaluation is not feasible using currently publicly-available datasets. Classification datasets either do not contain individual annotator decisions (only aggregated ones), do not include background information about the annotators, or are not robust enough to learn experts' behaviour in an online fashion. 
    %
    %
    While the focus of this paper is providing a feasible pipeline design and demonstrating its performance using theoretical analysis and multiple simulations, an important direction for future work on this framework would be evaluations using real-world online learning setups.
    }
    
    %
    %
    %
    %
    
    %
    \textbf{Exploration-exploitation transition.}
    Algorithms that employ other exploration-exploitation transition techniques (beyond Algorithm~\ref{alg:main_2}) can also be explored.
    In particular, prior works on multi-arm bandits and Thompson sampling provide a rich literature for methods that ensure a smooth transition from exploitation using prior information to exploration using incoming data streams \cite{russo2018tutorial}.
    However, applying these techniques
    in our setting can face similar challenges as those faced in the usage of Thompson sampling for contextual bandits \cite{riquelme2018deep}, and this direction can be additionally examined in future work on these algorithms.
    
    %
    \textbf{Biases and inaccuracies.} It is important to note that our framework aims to assign
    every expert to the subset of input space where they are expected to be accurate.
    For any expert, this characterization combines the subspaces where the expert is biased and subspaces where the expert is inaccurate, and aims to just find the complement of these spaces for this expert.
    %
    %
    This characterization is sufficient for our purposes since our goal is to make an accurate final prediction; however, in other settings, this characterization may be lacking.
    For example, an institution may want to invest resources towards explicitly addressing the shortcomings of the human experts.
    In this context, resources devoted towards addressing human biases would be different than those devoted towards addressing their fields of inexpertise; hence, separately learning these subspaces for any given human expert can be beneficial.
    %
    
\section{Conclusion}

We initiate a study of a hybrid pipeline where a classifier and multiple human experts share the decision-making load.
To train this pipeline, we provide algorithms that utilize the available human experts for labeling the training samples.
Even after the pipeline is sufficiently trained, inputs where the trained classifier confidence is low can still be deferred to the human experts, ensuring continual improvement and low error of final predictions.
%
%
Theoretical and empirical analysis shows that our algorithms can learn an accurate and unbiased pipeline, even when majority of the human experts are imperfect/biased.
With this pipeline, automated classifiers can be employed even when
training labels are unavailable but imperfect human experts are available for support.

When considered from the perspective of the entire decision-making pipeline, our framework 
takes into consideration the interaction between the training data labeling module and the optimization module.
A vast amount of literature in machine learning has been devoted to either learning algorithms that use labeled training data or guidelines for vigorous data collection processes.
Through our framework, we emphasize the importance of assessing and developing these two modules together, while focusing on handling human and data biases in a robust manner. 

\bibliographystyle{ACM-Reference-Format}
\bibliography{references}

\newpage
\appendix

\section{Details of \cite{keswani2021towards}} \label{sec:update_algorithm}
We employ the algorithm of \cite{keswani2021towards} as the subroutine in Algorithms~\ref{alg:main} and \ref{alg:main_2}.
For the sake of completion, we state the main (gradient-based) update step they propose here (\textbf{UpdateModel}).

Recall that $\mathcal{L}(u,v; \lambda, \alpha) = \mathcal{L}_f(u) + \alpha \mathcal{L}_D(u, v; \lambda),$
where $\mathcal{L}_f(u)$ is the classifier loss function, e.g., 
$$\mathcal{L}_f(u) :=  \E_{x,w,y} \left[ - y \log(f_u(x)) + (1-y)\log (1 - f_u(x))) \right],$$
and the deferrer loss function $\mathcal{L}_D(u, v; \lambda) =$
$$\E_{x,w,y}\left[ -y \log(\hat{y}_{u,v}) - (1-y)\log (1 - \hat{y}_{u,v}))  + \lambda \cdot D_u(x)^\top c(x) \right].$$
Hyperparameter $\alpha$ controls relative weight between classifier and deferrer training and $\lambda$ is the cost-hyperparameter.
%
%
%
\begin{algorithm}[!htbp]
\flushleft
  \caption{\textbf{UpdateModel} \cite{keswani2021towards, mozannar2020consistent} \\\textbf{Input}: training inputs $S = \set{(x_i, y_i)}_{i=1}^N$, classifier parameters $v$, deferrer parameter $u$, learning rate $\eta$, hyperparameters $\alpha, \lambda$.}
  \begin{algorithmic}[1] 
  	\State $u \gets u - \eta \pdv{\mathcal{L}(u,v; \lambda, \alpha)}{u}\Big{|}_{S}$
    \State $ \delta_v \gets $ projection of  $\pdv{\mathcal{L}(u,v; \lambda, \alpha)}{v}\Big{|}_{S}$ on simplex $\Delta^m$
    \State $v \gets v - \eta \cdot \delta_v$
	\State return $u,v$
\end{algorithmic}
    \label{alg:update}
\end{algorithm}
%
Note that other algorithms, such as \cite{mozannar2020consistent,madras2018predict} for the single-expert setting or MWU \cite{arora2012multiplicative} updates, can also be alternately be employed for updating the parameters given a batch of training inputs and their class labels.

\section{Proofs} \label{sec:proofs}


\textbf{Proof of Claim~\ref{clm:example}.}
Recall that $\alpha > 0.5$ fraction of experts are biased against group $z=0$ and $(1-\alpha)$ fraction are biased against group $z=1$; in other words, majority of the experts are biased against one group.
Assuming no prior, the training starts deferrer that assigns uniform weight $1/m$ to all experts. 
When $k=1$, the deferrer chooses a single expert to make the final decision.
Starting accuracy for group $z=1$ elements is $\alpha + 0.5(1-\alpha)$ and the starting accuracy for group $z=0$ elements is $(1-\alpha) + 0.5\alpha $. Hence the disparity between the two groups at step 0 is $(\alpha-0.5)$.

In the first training step, suppose we see a sample from group $z=0$ (will consider the other case later).
Training using prediction on this sample will have the following impact.
With probability $(1-\alpha)$ we will choose an unbiased expert and with probability $\alpha$ we will choose a biased expert.
The weight of the chosen expert is increased by quantity $\der$ and weight of other experts is decreased by quantity $\dep$; for appropriate normalization, $\delta := \der = (m-1)\dep$.
If an unbiased expert is chosen, the resulting accuracy for group $z=0$ (after the weights of all experts are updated) is 
\begin{align*}
    &\frac{\alpha m}{2} \left( \frac{1}{m} - \frac{\delta}{m-1} \right) + ((1-\alpha)m - 1) \left( \frac{1}{m} - \frac{\delta}{m-1} \right) + \frac{1}{m} + \delta\\
    =& \frac{\alpha m}{2} \left( \frac{1}{m} - \frac{\delta}{m-1} \right) + (1-\alpha) - (1-\alpha) \frac{m\delta}{m-1} + \frac{\delta}{m-1} \delta\\
    =& (1 - 0.5\alpha) + \frac{\delta\alpha m}{2(m-1)} .
\end{align*}
Since an unbiased expert is chosen, accuracy after one step of training increases by some amount.
If a biased expert is chosen instead, the resulting accuracy for group $z=0$ is 
\begin{align*}
    &(1-\alpha)m \left( \frac{1}{m} - \frac{\delta}{m-1} \right) + \frac{1}{2}\left( (\alpha m-1)\left( \frac{1}{m} - \frac{\delta}{m-1} \right) + \frac{1}{m} + \delta \right)\\
    =&(1-\alpha)m \left( \frac{1}{m} - \frac{\delta}{m-1} \right) + \frac{\alpha}{2} + \frac{\delta m}{2(m-1)} (1-\alpha)\\
    =& (1 - 0.5\alpha) - \frac{\delta}{2(m-1)} (1-\alpha)m.
\end{align*}
Hence, on expectation, the accuracy for group $z=0$ elements after one step of training (using a $z=0$ sample) is
\[(1 - 0.5\alpha) + (1-\alpha) \frac{\delta\alpha m}{2(m-1)} - \alpha \frac{\delta}{2(m-1)} (1-\alpha)m = (1-0.5\alpha).\]
Since $z=1$ accuracy will remain unchanged in this case, in expectation, the disparity between the accuracies for the two groups remains unchanged despite training.
In the above analysis, we did not use the fact that $\alpha > 0.5$. Hence the analysis for the case when we see a $z=1$ element is symmetric.

\textbf{Proof of claims in Remark~\ref{rem:example_dsim}.}
This time we start with a non-random deferrer, i.e., a deferrer induced by an appropriate $dSim$ function (i.e., for input $(x,z)$, we have that deferrer output $D(x)_i \propto dSim(e_i, z)$). 
Again we choose one expert to whom the decision is deferred.
Then,  for any input $x$ from group $z = 0$, the probability that we obtain the true label based on deferrer output is 
\[\frac{(1-\alpha) 1}{(1-\alpha) 1 + \alpha \gamma} + \frac{ \alpha \gamma}{(1-\alpha) 1 + \alpha \gamma} \cdot \frac{1}{2} = 1 - \frac{1}{2}\frac{\alpha \gamma}{(1-\alpha) 1 + \alpha \gamma}.\]
Similarly, for any input $x$ from group $z = 1$, the probability that we obtain the true label based on deferrer output is 
\[ 1 - \frac{1}{2}\frac{(1-\alpha) \gamma}{(1-\alpha) \gamma + \alpha 1}.\]
Hence, the starting disparity in accuracy for two groups is 
$$\dc := \frac{1}{2}\frac{\alpha \gamma}{(1-\alpha) 1 + \alpha \gamma} - \frac{1}{2}\frac{(1-\alpha) \gamma}{(1-\alpha) \gamma + \alpha 1}.$$
We can derive a lower bound on this quantity as follows
$$\dc \geq  \frac{1}{2}\frac{\alpha \gamma}{(1-\alpha)  + \alpha \gamma} + \frac{1}{2}\frac{(1-\alpha) \gamma}{(1-\alpha)  + \alpha \gamma}   = \frac{1}{2} \frac{\gamma}{1-\alpha + \alpha \gamma} \geq \frac{\gamma}{2}.$$
Similarly, we can also derive an upper bound as follows,
$$\dc \leq \frac{1}{2}\frac{\alpha \gamma}{(1-\alpha)  + \alpha \gamma}  \leq \frac{1}{2}\frac{\alpha \gamma}{(1-\alpha)}.$$
Note that $\alpha/(1-\alpha) > 1$ since $\alpha > 0.5$.
Hence,
\[\frac{\gamma}{2} \leq \dc \leq \frac{1}{2}\frac{\gamma}{(1-\alpha)}.\]
%
%
 %

\textbf{Proof of Theorem~\ref{thm:exploitation}.}
%
For input group $z$, let $d_i := dSim(e_i, z)$. 
By the condition in the theorem, for $j' \in \set{1, \dots, m} \setminus \set{j}$, we have that $d_{j'} < d_j - \beta$.
%
%
Then, if we select a single expert for deferral, the expected change in the weight assigned to expert $j$ is atleast
\[d_j \der - \left( \sum_{j' \neq j} (d_{j'} - \beta) \right) \dep.\]
The first term is the reward if expert $j$ is selected for making the decision while the second term is the penalty if some other expert is selected.
Since the overall updated weights have to be normalized, we have that $\delta := \der = (m-1) \dep$.
Therefore,
the expected change in the weight assigned to expert $j$ is atleast
\begin{align*}
d_j \delta - \left( \sum_{j' \neq j} (d_{j'} - \beta) \right) &\frac{\delta}{m-1}
= d_j \delta - \frac{\delta}{m-1} \sum_{j' \neq j} d_{j'} + \delta \beta \\
&= \frac{\delta}{m-1} \sum_{j' \neq j} (d_j - d_{j'}) + \delta \beta \geq 2\delta \beta.
\end{align*}

\textbf{Proof of Theorem~\ref{thm:exploration}.}
%
Probability that this expert is not selected in each of the $k$ samples is $(1-\epsilon)^k$. Since the total weight assigned by $dSim$ to accurate experts for group $z$ is greater than 0.5, this expert will be rewarded only if it is selected and otherwise penalized.
Therefore, the expected change in the weight of this expert is
\[(1-(1-\epsilon)^k) \der - (1-\epsilon)^k \dep. \]
On expectation, majority of experts in the chosen committee are correct, hence, due to the normalization constraint, we have that $\der = (m/k'-1)\dep$, where $k' > k/2$ are the number of correct experts.
Let $\epsilon' := (1-(1-\epsilon)^k)$.
For the expected change to be positive, we need
\begin{align*}
&\epsilon' \der > (1-\epsilon') \dep \\
\implies & \epsilon' \left(\frac{2m}{k} -1 \right) \dep > (1-\epsilon') \dep \\
\implies &  \frac{2m}{k} -1 > \frac{1}{\epsilon'}-1 \implies \epsilon' > \frac{k}{2m}.
\end{align*}
Substituting the value of $\epsilon'$, we get
\begin{align*}
(1-(1-\epsilon)^k) > \frac{k}{2m} &\implies  1-\epsilon < \left(1 - \frac{k}{2m} \right)^{1/k} \\
&\implies \epsilon > 1 - \left(1 - \frac{k}{2m} \right)^{1/k}.
\end{align*}

\section{Experimental details} \label{sec:experiments_appendix}

\subsection{Cluster detection task}
The classifier is a decision-tree model and the deferrer is a 2-layer neural network (with hidden layers of 16 and 8 nodes respectively). 

For Algorithm~\ref{alg:main}, the starting deferrer parameters are set in a manner such that the starting deferrer output simulates the $dSim$ function.
To do so, we first partition the set into two parts: the first part contains 500 elements and second part contains 1500 elements.
The first unlabeled part is used to set the initial deferrer parameters by regressing the network on the inputs and the corresponding $dSim$ values.
We use stochastic gradient descent over mean-squared error loss function for this training, with a learning rate of 0.001 and 500 training steps.

%
For the main training of Algorithm~\ref{alg:main}, we
follow the gradient descent approach of \cite{keswani2021towards}, using the predicted class label from the framework as the true label. The learning rate is 0.0075 and the batch size is kept to 10.

\subsection{Content moderation task} \label{sec:content_moderation_appendix}

For the content moderation task, we obtain the dialect of the posts using the dialect identification model provided by \citet{blodgett2017dataset}.

\textbf{Implementation Details. }
The dataset here is split into train and test partitions (80-20 split); with the first partition used as a stream for online training and second partition used for testing.
Both classifier and deferrer are neural networks with three hidden layers with ReLU activation.

Once again, we use the algorithm of \cite{keswani2021towards} (Appendix~\ref{sec:update_algorithm}) as a subroutine.
The experts are given a cost of 1 each, i.e., $c_i(x) = 1$, for all $i \in \set{1, \dots, m-1}$, and $c_m(x) =0$.
The cost hyperparameter $\lambda = t/100$, where $t$ is the iteration number (giving higher cost to classifier training in initial iterations).

%
%
The learning rate $\eta=0.01$
%
with batch size of $B=100$ per iteration.

%

For Algorithm~\ref{alg:main}, once again we assume a small number (1000) of initial unlabeled samples are provided for training the initial deferrer.
This initial regression training to obtain a starting deferrer uses the Adam \cite{kingma2014adam} algorithm over mean-square loss with learning rate 0.0001 and 1000 iterations.
For Algorithm~\ref{alg:main_2}, we set the parameter $T_d = 10000$.

\begin{figure}
    \centering
    \includegraphics[width=\linewidth]{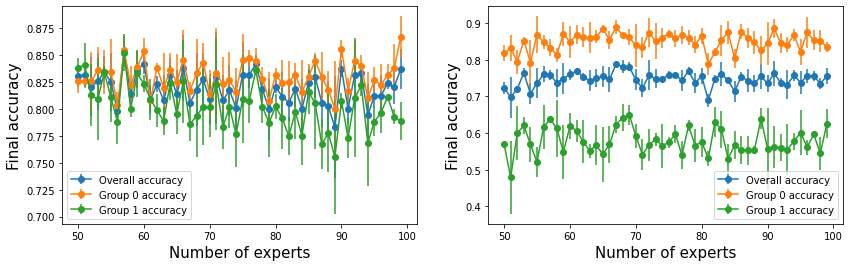}
    \subfloat[Performance of \ref{alg:main}]{\hspace{.5\linewidth}}
    \subfloat[Performance of \ref{alg:main_2}]{\hspace{.5\linewidth}} 
    \caption{Performance of our approaches for different number of experts in the content moderation task.}
    \label{fig:acc_vs_nexperts}
\end{figure}

\textbf{Figure~\ref{fig:acc_vs_nexperts}} finally presents the variation of accuracy of both algorithms with number of available experts.

\end{document}